\documentclass[12pt,preprint]{aastex}

\newcommand{\fracp}[2]{\frac{\partial #1}{\partial #2}}

\def\msun{{M_\odot}}
\def\kms{\,{\rm {km\, s^{-1}}}}                                             
\def\pppm{\rm P^3M}
\def\mpc{\,h^{-1}{\rm {Mpc}}}

\shorttitle{SPATIAL AND DYNAMICAL BIASES IN VELOCITY STATISTICS}
\shortauthors{YOSHIKAWA ET AL.}

\begin{document}

\title{SPATIAL AND DYNAMICAL BIASES IN VELOCITY STATISTICS OF GALAXIES}

\author{Kohji Yoshikawa} 
\affil{Research Center for the Early Universe (RESCEU), School of Science, University of Tokyo, Tokyo 113-0033, Japan.}
\email{kohji@utap.phys.s.u-tokyo.ac.jp}

\author{Y.P. Jing} 
\affil{Shanghai Astronomical Observatory, the Partner Group of MPI f\"ur
Astrophysik, \\Nandan Road 80,  Shanghai 200030, China}
\email{ypjing@center.shao.ac.cn}

\and

\author {Gerhard B\"orner}
\affil {Max-Planck-Institut f\"ur Astrophysik,
Karl-Schwarzschild-Strasse 1, \\  85748 Garching, Germany}
\email{grb@mpa-garching.mpg.de}

\received{2002 ???}
\accepted{2003 ???}

\begin{abstract}
 We present velocity statistics of galaxies and their biases inferred
 from the statistics of the underlying dark matter using a cosmological
 hydrodynamic simulation of galaxy formation in low-density and
 spatially flat ($\Omega_0=0.3$ and $\lambda_0=0.7$) cold dark matter
 cosmogony. Our simulation is based on a $\pppm$ $N$-body Poisson solver
 and smoothed particle hydrodynamics. Galaxies in our simulation are
 identified as clumps of cold and dense gas particles, and classified as
 young and old galaxies according to their formation epochs. We find
 that the pairwise velocity dispersion (PVD) of all galaxies is
 significantly lower than that of the dark matter particles, and that
 the PVD of the young galaxies is lower than that of the old types, and
 even of all galaxies together, especially at small separations. These
 results are in reasonable agreement with the recent measurements of
 PVDs in the Las Campanas redshift survey, the PSC$z$ catalogue and the
 SDSS data. We also find that the low PVD of young galaxies is due to
 the effects of dynamical friction as well as the different spatial
 distribution, while the difference in the PVD between all galaxies and
 dark matter can be mostly ascribed to their different spatial
 distributions.  We also consider the mean infall velocity and the
 POTENT density reconstruction that are often used to measure the
 cosmological parameters, and investigate the effects of spatial bias
 and dynamical friction. In our simulation, the mean infall velocity of
 young galaxies is significantly lower than that of all the galaxies or of
 the old galaxies, and the dynamical bias becomes important on scales
 less than $3\mpc$. The mass density field reconstructed from the
 velocity field of young galaxies using the POTENT-style method suffers
 in accuracy both from the spatial bias and the dynamical friction on
 the smoothing scale of $R_s=8\mpc$.  On the other hand, in the case of
 $R_s=12\mpc$, which is typically adopted in the actual POTENT analysis,
 the density reconstruction based on various tracers of galaxies is
 reasonably accurate. We also analyze the motions of central galaxies
 and the velocity dispersion of galaxies within halos, and discuss their
 implications for the motion of cD galaxies and the determination for
 the mass of galaxy groups.
\end{abstract} 


\section{INTRODUCTION}
It has been widely attempted to model the dark matter distribution in
the universe using the velocity information of galaxies, based on the
assumption that the velocity is induced by the gravity. The peculiar
velocity of galaxies can either be measured from secondary distance
indicators (e.g. the Tully-Fisher relation for spirals and the
fundamental plane for ellipticals) or it can be inferred from the
distortion of the galaxy distribution in redshift space. Two simple
statistics are often used to describe the velocity of galaxies; the
mean infall velocity $v_{12}(r)$ and the pairwise velocity dispersion
$\sigma_{12}(r)$ (hereafter PVD). These statistics are extensively
used to constrain the density parameter $\Omega_0$ of the universe and
the clustering of dark matter on small scales \citep{dp83,
Davis1985,os90,Mo1993, Fisher1994,sutojing97,Jing1998,
Juszkiewicz2000, Jing2002, Zehavi2002}. In addition, the observed
peculiar velocity of galaxies has been used to measure the density
field of the nearby universe, and the parameter
$\beta\equiv\Omega_0^{0.6}/b$ by a number of authors
\citep{Bertschinger1989, Dekel1994, Nusser1994, Sigad1998, Dekel1999,
Peacock2001, Zaroubi2002}. Here $b$ is the linear bias factor of
galaxy clustering.

Using the Las Campanas Redshift Survey (LCRS) of galaxies
\citep{Shectman1996}, \citet{Jing1998} analyzed the PVD of galaxies and
measured $\sigma_{12}=580\pm 80 \kms$ at the projected separation
$r_{\rm p}=1\mpc$. The uncertainty quoted was obtained from an analysis
of a set of mock samples. Since the LCRS contains about a few tens of
clusters of galaxies, the PVD can be determined from this survey to an
accuracy of 15 percent (including the cosmic variance). The result has
been recently confirmed by \citet{Zehavi2002} and \citet{Hawkins2003}
who measured the PVD of galaxies for the Early Data Release of the Sloan
Digital Sky Survey (SDSS) and the 2dF Galaxy Redshift Survey (2dFGRS),
respectively. In addition, it has been found that the PVD of late-type
galaxies is smaller than that of early-type ones \citep{Mo1993,
Fisher1994}. \citet{Jing2002} extensively analyzed the PVD of the
galaxies in the PSC$z$ catalogue \citep{Saunders2000}, which are mostly
late-type galaxies, and found that the PVD is about $\sim 400\kms$ at
$r_{\rm p}=2\mpc$, and rapidly decreases to $\sim 200 \kms$ at small
separation $r_{\rm p}=0.2\mpc$. This finding for the {\it IRAS} galaxies
is again in good agreement with the SDSS result for blue or late-type
galaxies by \citet{Zehavi2002}. As for the PVD of dark matter, a number
of previous numerical works \citep[e.g., ][]{Benson2000, Suto1993,
Jenkins1998, Jing1998, Jing2002} conclude that the PVD of dark matter is
$\sigma_{12}\simeq700-800\kms$ at $r_{\rm p}=1\mpc$ for the
spatially-flat low-density CDM universe, and systematically higher than
those of observed galaxies.

There are two causes for the difference of the velocity dispersion of
galaxies and of dark matter. One is the fact that galaxies are
distributed spatially differently from dark matter particles. Indeed,
\citet{Jing1998} proposed a cluster (under)weighted (CLW) model for
explaining both the spatial two-point correlation function and the
pairwise velocity dispersion measured in the LCRS in popular flat
low-density cold dark matter (LCDM) models. In this model, it is assumed
that the number of galaxies per unit mass is a decreasing function of
the hosting cluster (halo) mass, i.e. galaxies are anti-biased in rich
cluster regions (see \citet{Carlberg1996} and \citet{Bahcall2002} for
the observational evidence for the anti-biasing). The resulting
difference in the PVDs between galaxies and dark matter is purely due to
the difference between the two populations in their spatial
distribution. We will call this bias in the velocity statistics the
``spatial bias''.  Assuming that the {\it IRAS} galaxies are more
strongly anti-biased in rich cluster regions than the optical galaxies,
\citet{Jing2002} found that the spatial clustering and the PVD of {\it
IRAS} galaxies can be quantitatively explained in the popular LCDM
models, except for the rapid decrease of the PVD at small separations
observed for the {\it IRAS} galaxies.

Another cause for the velocity bias is dynamical friction. Due to this
mechanism, the motion of galaxies could be systematically slower than
that of dark matter particles within a host halo. We will call this bias
the ``dynamical bias''. There were many studies to quantify dynamical
biases, but no consensus has been reached as to which extent the
velocity statistics are affected by the dynamical bias \citep{Katz92,
Frenk96, Cen00, Pearce01}. In \citet{Jing2002}, it was conjectured that
the rapid decrease of the PVD of {\it IRAS} galaxies is a result of the
dynamical bias. Since a similar decrease of the PVD was not found for
optical galaxies, a systematic study is needed to quantify how the
velocity statistics of different populations of galaxies are changed by
the dynamical bias.

In this paper, we will use a cosmological hydrodynamic simulation to
study the effect of spatial and dynamical biases on several velocity
statistics of galaxies. Galaxies are identified in the simulation as
clumps of cold and dense gas, and classified into blue and red
populations based on their formation epochs. We analyze the velocity
statistics for different populations of galaxies as well as for dark
matter. We measure not only the PVDs of the different tracers, but also
consider the mean infall velocity and the POTENT-style analysis. The
later two analyses have been used to measure the dynamical parameter
$\beta=\Omega_0^{0.6}/b$ based on the (quasi-)linear perturbation theory
of dark matter clustering \citep{Juszkiewicz2000, Dekel1994,
Nusser1994}. Since the observed velocity fields consist of the velocity
data of spirals and/or ellipticals, it is important to quantify whether
the velocity statistics calculated based on various galaxy populations
are biased relative to those of dark matter.

We will separate the effects of the spatial and dynamical biases on
these velocity statistics by constructing dynamically unbiased mock
samples of galaxies and comparing their velocity statistics with those
of galaxies in our simulation. One would expect that the dynamical
friction mechanism operates in small dense regions, i.e. on small
scales, but it is important to quantify on which scale the dynamical
bias is significant for the statistics considered. The reason to
separate the two biases is that the spatial bias is relatively easy to
model for the velocity statistics if the occupation numbers of
galaxies within halos are known or constrained from other observations
\citep{Mo1997, Jing1998, Seljak00, ps01,Sheth01a, Sheth01b, ss02,
Berlind02, Zheng02, Kang02, Kuwabara02, Yang02}. We will show
quantitatively where the spatial bias models are applicable in
interpreting the observations.  In addition, in order to see the
effect of dynamical friction on galaxies in more direct manner, we
also examine the motions of individual galaxies relative to their host
dark halos, and the relation between the velocity dispersion of member
galaxies and that of dark matter in dark halos.

The rest of the paper is organized as follows. In \S 2, we briefly
describe our numerical simulation and our procedure to identify and
classify galaxies. In \S 3, some statistics for the velocity field of
galaxies and dark matter are analyzed with emphasis on the effects of
the spatial and dynamical biases. Finally, we summarize our findings in
\S 4.

\section{SIMULATION}
Our simulation is the same as that described in \citet{Yoshikawa2001},
and the details about the simulation can be found in
\citet{Yoshikawa2000} and \citet{Yoshikawa2001}. Here, we briefly
summarize some basic features.

The simulation code is a hybrid of a Particle--Particle--Particle--Mesh
($\pppm$) Poisson solver and smoothed particle hydrodynamics (SPH). We
adopt $128^3$ dark matter particles and the same number of gas particles
within a periodic simulation box of $L_{\rm box} =75 \mpc$ per side. The
cosmological model considered in this paper is a spatially-flat
low-density CDM universe with $\Omega_0=0.3$, $\lambda_0=0.7$,
$\sigma_8=1.0$, and $h=0.7$, where $\lambda_0$ is the dimensionless
cosmological constant, $\sigma_8$ the rms density fluctuation at a scale
of $8\mpc$, and $h$ the Hubble parameter in units of 100
km/s/Mpc. Furthermore, we set the baryonic density parameter to
$\Omega_{\rm b}=0.015h^{-2}$ \citep{Copi1995}. Therefore, the mass of a
single gas and dark matter particle is $2.4\times 10^{9} M_{\odot}$ and
$2.2\times 10^{10} M_{\odot}$, respectively. The ideal gas equation of
state with an adiabatic index $\gamma=5/3$ is adopted, and we consider
the effect of radiative cooling adopting the cooling rate with the
metallicity [Fe/H]$=-0.5$ from \citet{Sutherland1993}. In order to avoid
numerical overcooling of gas particles in SPH scheme, we implement the
cold gas decoupling scheme which is first introduced by
\citet{Pearce1999} (see \citet{Yoshikawa2001} for the detail
implementation). The initial condition is created at redshift $z=36$ and
is evolved to $z=0$.

We have 50 outputs at different redshifts between $z=9$ and $z=0$. For
each output, galaxies are identified using the following
procedure. First, we extract gas particles which satisfy the Jeans
condition
\begin{equation}
\label{eq:jeans}
h_{\rm\scriptscriptstyle SPH} > \frac{c_s}{\sqrt{\pi G \rho}},
\end{equation}
and the overdensity condition
\begin{equation}
\label{eq:density}
\rho > 10^3\bar{\rho}_{\rm b}(z),
\end{equation}
where $h_{\rm\scriptscriptstyle SPH}$ is the smoothing length of gas
particles, $c_s$ the sound speed, $G$ the gravitational constant, $\rho$
the SPH gas density, and $\bar{\rho}_{\rm b}(z)$ the cosmic mean baryon
density at redshift $z$. Then, we use the friend-of-friend (FOF)
algorithm \citep{Davis1985} to identify galaxies as clumps of these
Jeans unstable gas particles. Here, we adopt the FOF linking length of
$b=0.0164(1+z)\bar{l}$ as used in \citet{Yoshikawa2001}, where $\bar{l}$
is the mean separation of gas particles. We consider only galaxies with
baryonic mass greater than $10^{11}\msun$, which is equivalently to 40
times of the gas particle mass.

\begin{figure}[tbp]
\leavevmode
\begin{center}
 \plotone{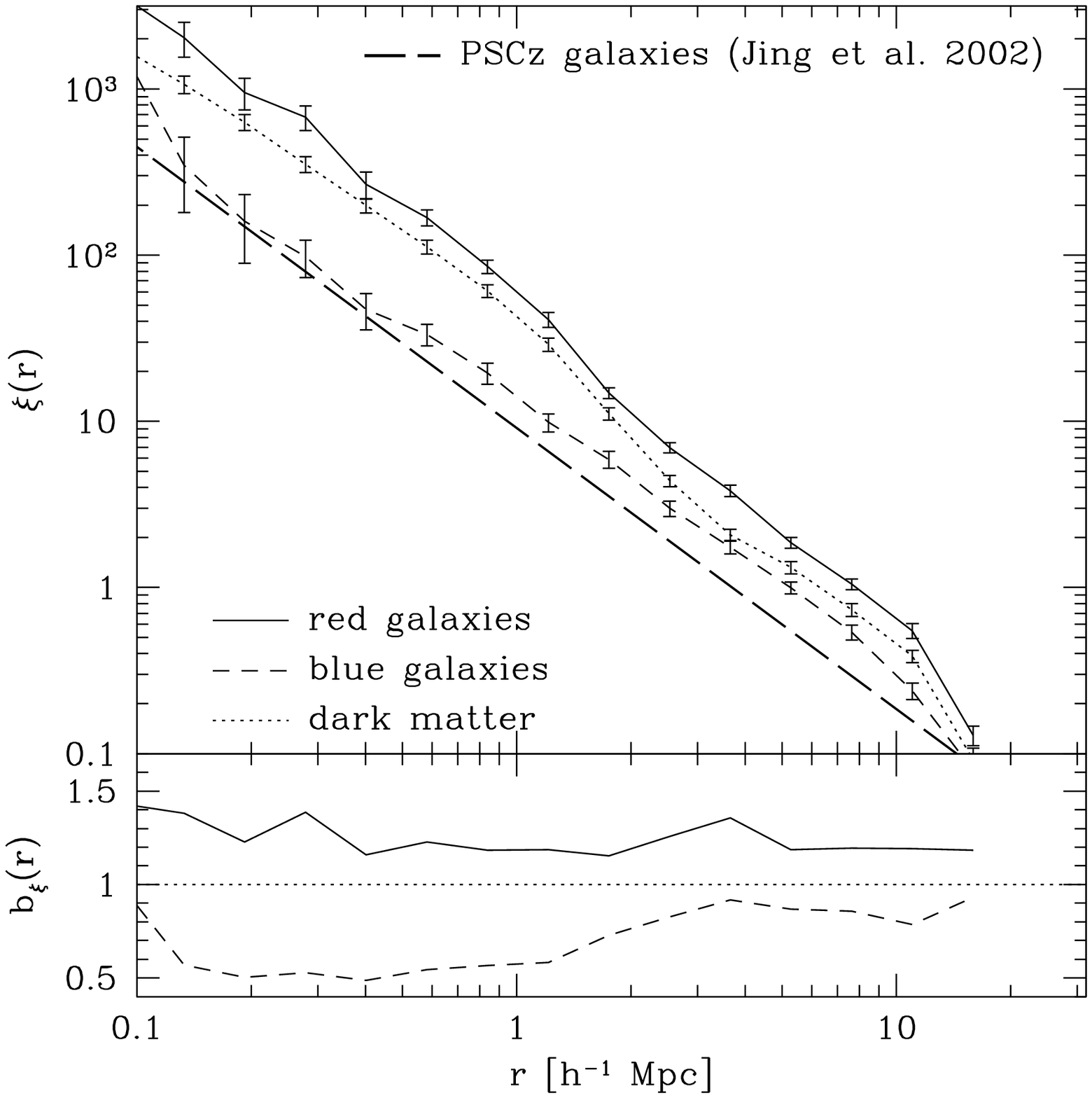}
\figcaption{The two-point correlation functions for blue galaxies, red
galaxies, and dark matter are shown in the {\it upper} panel. The
profiles of the bias parameter $b_{\xi}(r)$ are also shown in the {\it
lower} panel. For reference, we plot the two-point correlation
function for the PSC$z$ galaxies derived by
{\protect \citet{Jing2002}}\label{fig:xi}}
\end{center}
\end{figure}

For each galaxy identified at $z=0$, we define its formation redshift
$z_{\rm f}$ as the epoch when half of its member particles satisfy the
conditions (\ref{eq:jeans}) and (\ref{eq:density}). 
Using the formation redshift
$z_{\rm f}$ of simulated galaxies, we classify those galaxies at $z=0$
into two groups, the ones with $z_{\rm f}<1.8$ and the others with
$z_{\rm f}>1.8$, which we call, for later convenience, the {\it blue}
galaxies and the {\it red} galaxies, respectively. The value of threshold
redshift $z_{\rm f}=1.8$ is chosen so that the blue and red population
have almost same number of galaxies just for the statistical
significance of later analyses. In the procedures described above, we
finally identify 1062 blue galaxies, 962 red galaxies, and 2024 galaxies
in total at $z=0$. Here, it is not expected that such a simple
categorization can explain the observed properties of early- and
late-types of galaxies in any detail, but it can reproduce some
statistical properties of observed galaxies qualitatively. Actually, as
already discussed in \citet{Yoshikawa2001}, the red galaxies reside
within massive dark halos, and the blue ones are preferentially formed
in smaller dark halos. In addition, considering the observed correlation
between galaxy types and its formation histories, blue and red galaxies
in our simulation roughly correspond to late- and early-types of
galaxies, respectively. Figure~\ref{fig:xi} shows the two-point
correlation functions for red and blue galaxies. The profiles of the
biasing parameter for each population of galaxies defined by
\begin{equation}
b_{\xi}(r) = \sqrt{\frac{\xi_{\rm gg}(r)}{\xi(r)}},
\end{equation}
are also shown in the lower panel, where $\xi_{\rm gg}(r)$ and $\xi(r)$
are two-point correlation functions for galaxies and dark matter,
respectively. The error bars indicate the 1-$\sigma$ poisson error. We
can see that the red galaxies exhibit a relatively higher amplitude of
two-point correlation functions than the blue ones. This feature is
consistent with the observed morphology dependent clustering of galaxies
\citep[e.g.,][]{Loveday1995, Hermit1996, Norberg2002, Zehavi2002}. We
also plot the observed two-point correlation function of galaxies in the
PSC$z$ catalogue,
\begin{equation}
\xi_{\rm PSC\it z}(r)=\left(\frac{r}{3.7\mpc}\right)^{-1.69}
\end{equation}
estimated by \citet{Jing2002}. The two-point correlation function of the
blue galaxies is relatively close to that of the PSC$z$ galaxy sample
compared with all and red galaxies. It is amazing to note that the
$\xi(r)$ of the blue galaxies in our simulation is very close to the
prediction of the cluster-weighted model of $\alpha=0.25$ in
\citet{Jing2002} (see their Fig.5). This spatial bias model with
$\alpha=0.25$ has been shown in good agreement with the projected
two-point correlation function estimated from the PSC$z$ catalogue. This
means that the blue galaxies in our simulation well represent the
clustering properties of the galaxies in the PSC$z$ catalogue.
Furthermore, since galaxies in the PSC$z$ catalogue are selected in the
infrared waveband, they presumably have experienced star formation very
recently, or they are just undergoing star formation.  Therefore, it is
quite natural to consider that the blue galaxies in our simulation,
which are formed recently, indeed represent the physical properties of
the PSC$z$ galaxies.

\section{VELOCITY STATISTICS}
\subsection{PAIRWISE VELOCITY DISPERSION}
Observationally, PVDs of galaxies can be measured by modeling the
distortion of their two-point correlation function in
redshift-space. Therefore, observed PVDs are inevitably projected along
the line of sight and functions of the projected separation. While, of
course, three dimensional PVDs obtained from the simulation cannot be
directly compared with the observed ones, according to
\citet{Jenkins1998}, we can calculate the projected line-of-sight PVD
using three dimensional PVD as
\begin{equation}
 \label{eq:PVD_projected}
 \sigma^2_{12}(r_{\rm p}) = \frac{\displaystyle \int \xi(r)\sigma^2_{\rm proj}(r)dl}{\displaystyle \int \xi(r)dl},
\end{equation}
where $\xi(r)$ is the two-point correlation function of objects we are
considering, $r_{\rm p}$ and $l$ denotes the projected separation and
the distance along the line-of-sight, respectively, and therefore $r^2 =
r_{\rm p}^2 + l^2$. The quantity $\sigma_{\rm proj}^2$ is given by
\begin{equation}
 \sigma^2_{\rm proj} = \frac{r_{\rm p}^2\sigma_{\rm v}^2+l^2(\sigma_{\perp}^2-v_{12}^2)}{r_{\rm p}^2+l^2},
\end{equation}
where $\sigma_{\perp}^2$ is the velocity dispersion perpendicular to the
vector connecting each pair, and $v_{12}$ is the mean infall velocity
described below.

\begin{figure}[tbp]
 \leavevmode
 \begin{center}
  \plotone{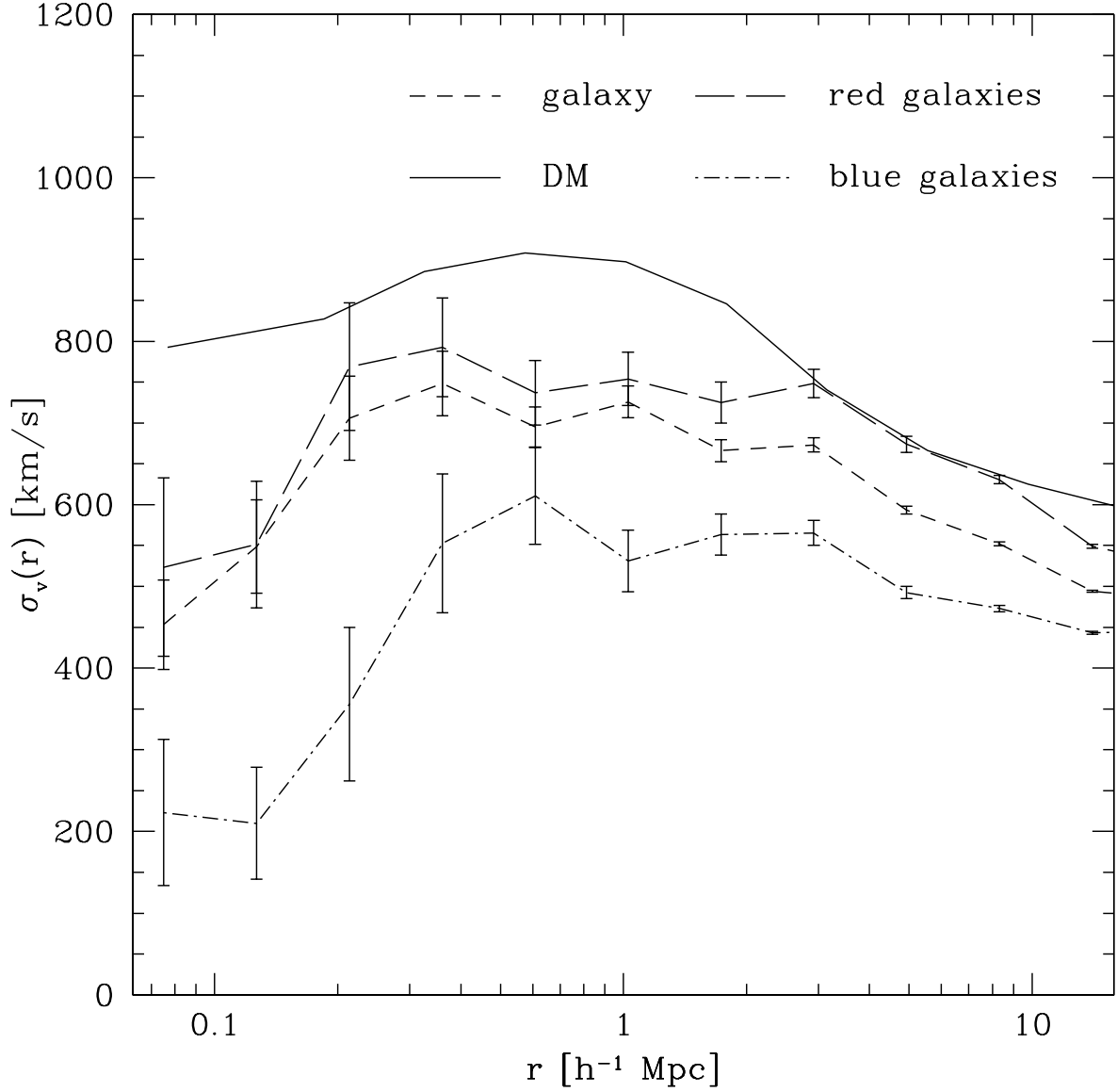}
  \figcaption{The three-dimensional pairwise velocity dispersions of
  dark matter particles, all galaxies, red and blue
  galaxies. \label{fig:PVD}}
 \end{center}
\end{figure}

\begin{figure}[tbp]
 \leavevmode
 \begin{center}
  \plotone{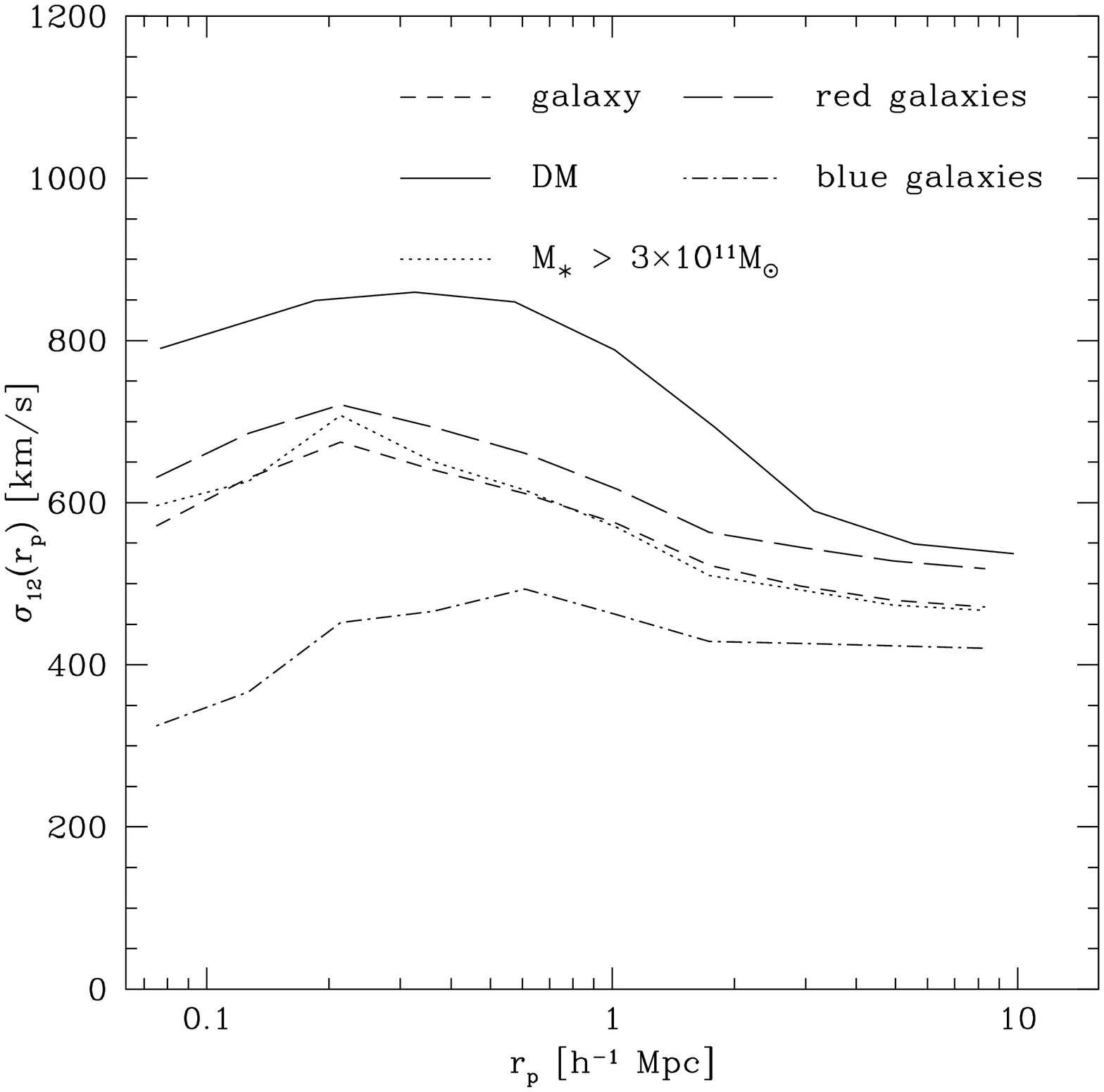}
  \figcaption{Same as Fig.~\ref{fig:PVD} but for the PVDs projected
  along line-of-sight. \label{fig:PVD_projected}}
 \end{center}
\end{figure}

Figure~\ref{fig:PVD} shows the three dimensional PVDs of dark matter and
galaxies in our simulation. The quoted error bars indicate the
1-$\sigma$ poisson error. As is shown in previous works \citep{Jing1998,
Benson2000}, the PVD of all galaxies is markedly lower than that of the
dark matter particles on all the scales considered. In
Figure~\ref{fig:PVD}, we also show the PVDs of blue and red galaxies,
separately. In Figure~\ref{fig:PVD_projected} we show the projected
line-of-sight PVDs of dark matter, all galaxies, blue and red
galaxies. One can see that the PVD of the blue galaxies is much lower
than that of all the galaxies or of the red population, especially at
small separation, down to $\sigma_{\rm v}\sim200\kms$ at $r=0.1\mpc$ and
$\sigma_{12}\sim350\kms$ at $r_{\rm p}=0.1\mpc$. The PVD of the red
galaxies is almost the same as or slightly higher than that of all
galaxies. This behavior of the PVDs of the three populations of galaxies
is qualitatively consistent with the PVD of galaxies in the LCRS
\citep{Jing1998}, the PSC$z$ \citep{Jing2002} and also the SDSS data
\citep{Zehavi2002}. In order to see the effect of selection of galaxies,
in Figure~\ref{fig:PVD_projected}, we also show the PVD of galaxies with
higher mass $M_{\star}>3\times10^{11}M_{\odot}$. One can see that it is
almost the same as the PVD of all galaxies, and that our estimate of PVD
is not sensitive to the selection of galaxies. The PVDs in the 2dFGRS
\citep{Hawkins2003} and the LCRS \citep{Jing1998} are in good agreement
with our result for large separation $r_{\rm p}\gtrsim0.5\mpc$, though
our simulated PVD is slightly higher at $r_{\rm p}<0.5\mpc$. The
difference at small separations can be partially ascribed to the effect
of cosmic variance, because PVD is generally quite sensitive to the
presence and/or absence of rich galaxy clusters, and because our
simulation volume is smaller than the survey volumes of the 2dFGRS and
the LCRS.

\begin{figure}[tbp]
 \leavevmode 
 \begin{center}
  \plotone{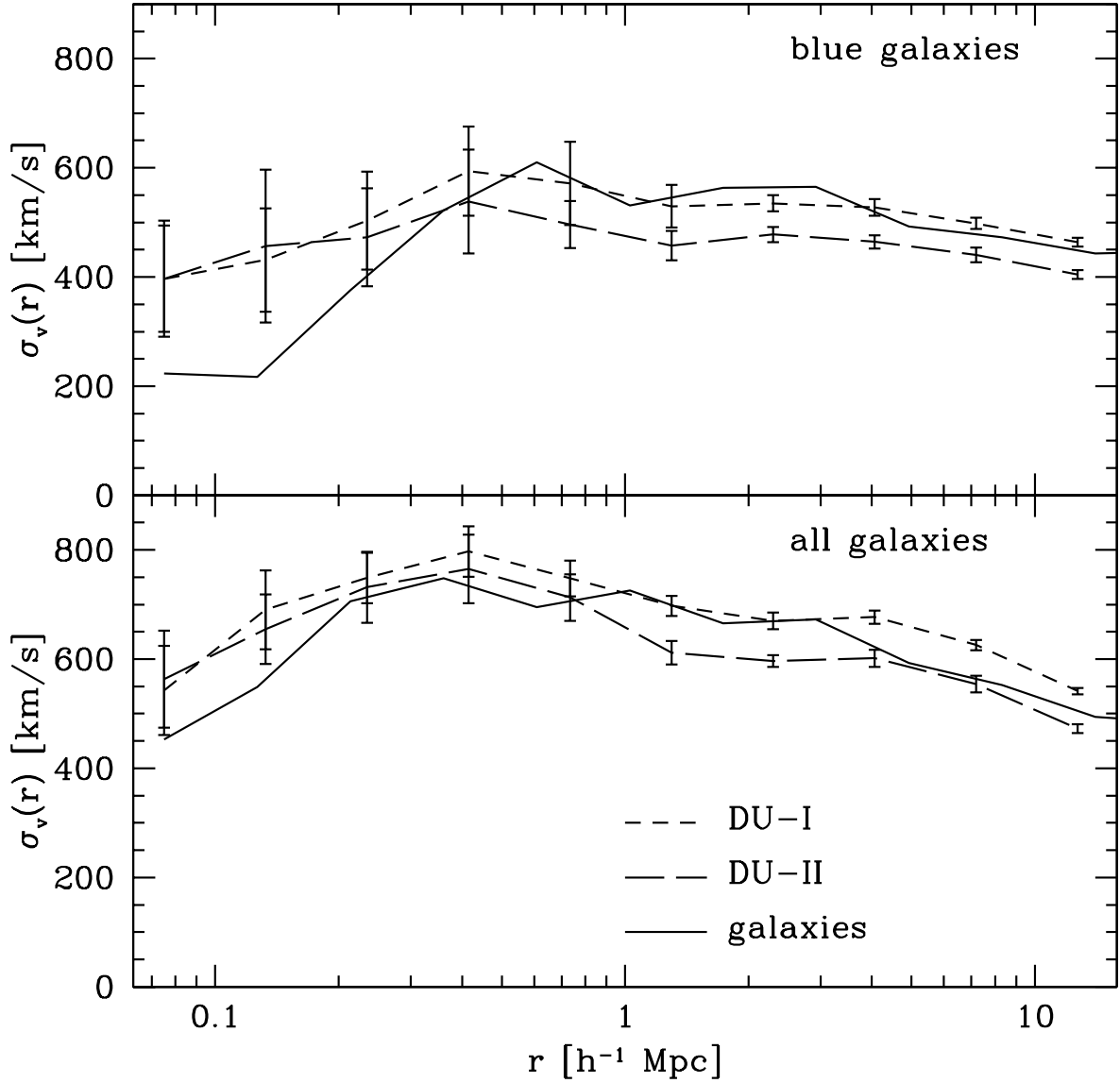}
 \figcaption{The pairwise velocity dispersions of the two dynamically
 unbiased samples and galaxies for the whole population ({\it lower}
 panel) and the blue population ({\it upper} panel).
 \label{fig:mockPVD}} \end{center}
\end{figure}

\begin{figure}[tbp]
 \leavevmode 
 \begin{center}
  \plotone{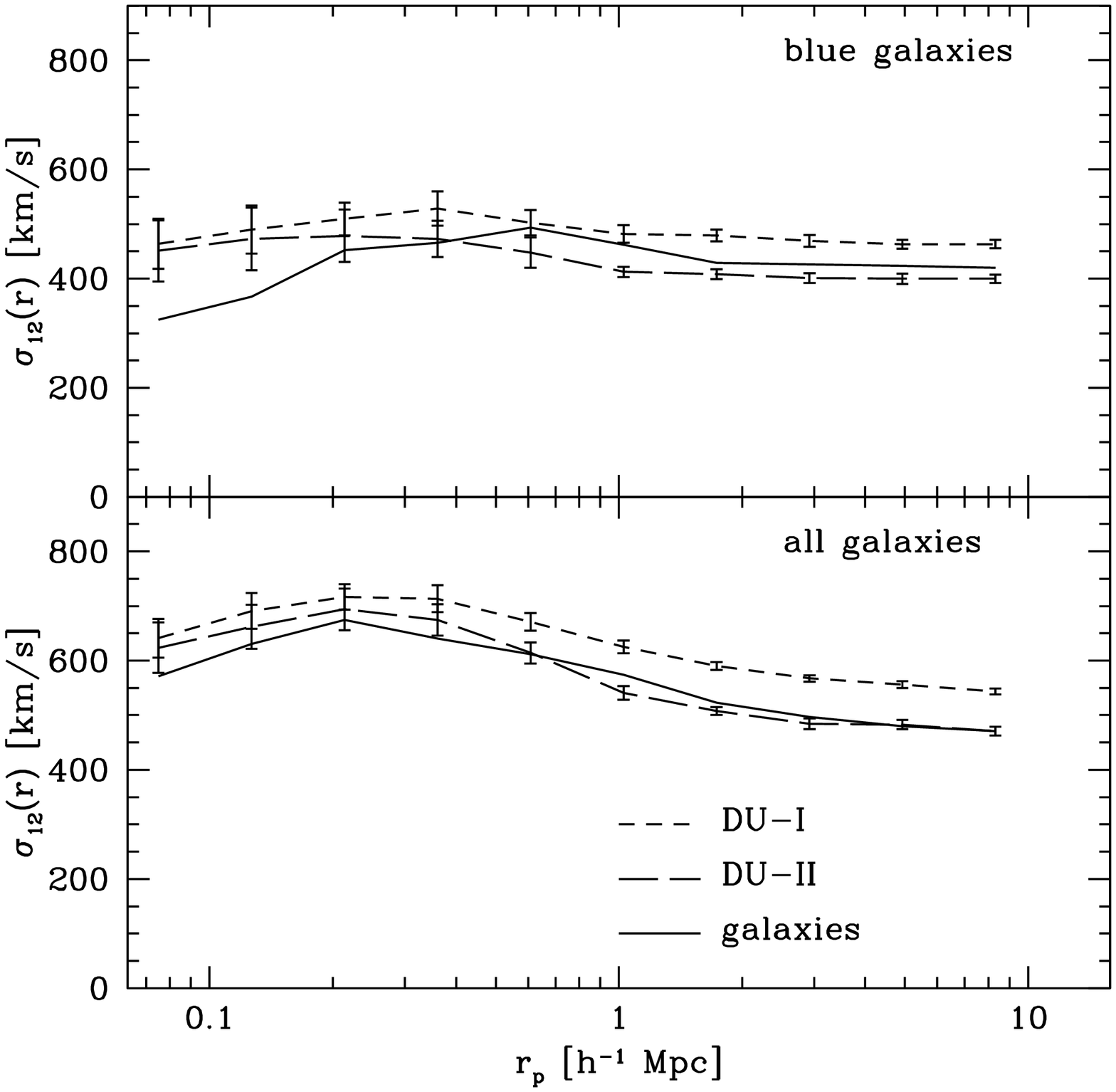}
  \figcaption{Same as Fig.~\ref{fig:mockPVD} but for the PVDs projected
  along line-of-sight.
 \label{fig:mockPVD_projected}} \end{center}
\end{figure}

In order to discriminate between the effects of spatial and dynamical
biases on the PVDs of galaxies, we create two types of mock galaxy
samples from our simulation data. In the first sample, for each galaxy,
we randomly select one {\it dark matter} particle within its host
dark halo, and assign to it the same formation redshift.  The second
sample is the same as the first one, except that the particles
corresponding to the central massive galaxies in each halo are assigned
the bulk motion velocity of the host halo, i.e. these particles are
static in their host dark halos. Therefore, these samples mimic the
spatial distribution of galaxies but keep the properties of the dark
matter velocity field, and can be regarded as spatially biased and
dynamically unbiased samples. For later convenience, we call the first
and the second dynamically unbiased samples DU-I and DU-II
respectively. We construct 20 realizations of these dynamically unbiased
samples of all galaxies and of the blue galaxies by selecting particles
with the assigned formation redshift less than 1.8. We confirm that the
two-point correlation functions of these samples agree well with those
of corresponding galaxy samples. In Figure~\ref{fig:mockPVD}, we show
the three dimensional PVDs and of the DU-I and DU-II samples for all the
galaxies as well as for the blue galaxies. The error bars indicate the
statistical dispersion for 20 realizations of the samples. One can see
that the PVD of all galaxies (solid line in the lower panel) is in
reasonable agreement with DU-I and DU-II samples. This indicates that
the lower PVDs of all the galaxies can be explained by the different
spatial distribution between galaxies and dark matter. This result is
consistent with previous work \citep{Jing1998, Benson2000} and we
confirm it more directly by comparing the galaxy sample and the
dynamically unbiased samples. On the other hand, the PVD of the blue
galaxies (solid line in the upper panel) deviates downward from that of
the DU-I and DU-II samples at small separation of $r < 0.3\mpc$,
therefore it cannot be explained only by the spatial bias.
Figure~\ref{fig:mockPVD_projected} shows that the line-of-sight
projected PVDs exhibit the same results as in the case of the
three-dimensional PVDs. We can suggest that that the rapid decrease of
the PVD of late-type galaxies found in the PSC$z$ catalogue
\citep{Jing2002} and SDSS galaxies \citep{Zehavi2002} is due to the
effect of dynamical friction. More evidence will be given in
\S~\ref{ss:motion_of_galaxies}.

Let us note the comparison with the previous works by \citet{Benson2000}
and \citet{Kauffmann1999}. First, it should be noted that in these two
works, ``galaxies'' are formed according to combination of
dissipationless $N$-body simulations and semi-analytic galaxy formation
schemes, while the ``galaxies'' in our SPH simulation are identified as
clumps of cold and dense gas particles. \citet{Kauffmann1999} found very
similar PVDs for galaxies and dark matter, which are not consistent with
recent observational results \citep{Jing1998, Jing2002, Zehavi2002}, and
\citet{Benson2000}. Actually, the line-of-sight projected PVD of all
galaxies in our simulation also significantly departs from that of dark
matter, and is consistent with the result of \citet{Benson2000}.
\citet{Benson2000} argued that the discrepancy with
\citet{Kauffmann1999} is due to the different frequency with which dark
halos are populated with a particular number of galaxies in their
semi-analytic schemes. Since our galaxies are identified in totally
different manner from \citet{Benson2000} and \citet{Kauffmann1999} as
described above, our result gives independent verification for the
result by \citet{Benson2000}.  In \citet{Benson2000}, since the
galaxies' positions and velocities are assigned those of dark matter
particles which are randomly chosen within host dark matter halos,
except for central galaxies of dark halos, which are assigned the
positions and velocities of mass centers of their host dark halos, their
galaxy sample is, by construction, free from the dynamical friction, and
somewhat similar to our DU-II sample. Therefore, their estimation of
galaxies' PVD missed the effect of dynamical bias which we focus on
throughout in this paper.

\subsection{MEAN INFALL VELOCITY}

\begin{figure}[tbp]
 \leavevmode 
 \begin{center}
  \plotone{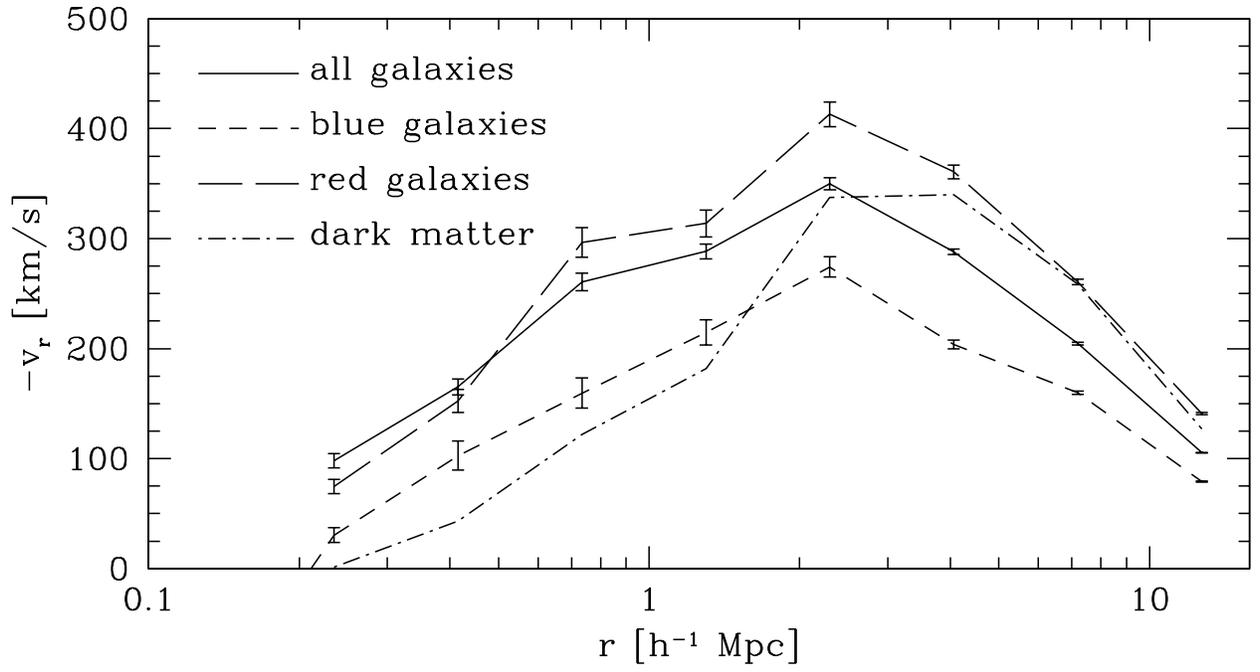}
 \end{center} \figcaption{The mean infall velocity of dark matter, all
 galaxies, blue galaxies, and red galaxies.\label{fig:vinfall}}
\end{figure}

\begin{figure}[tbp]
 \leavevmode 
 \begin{center}
  \plotone{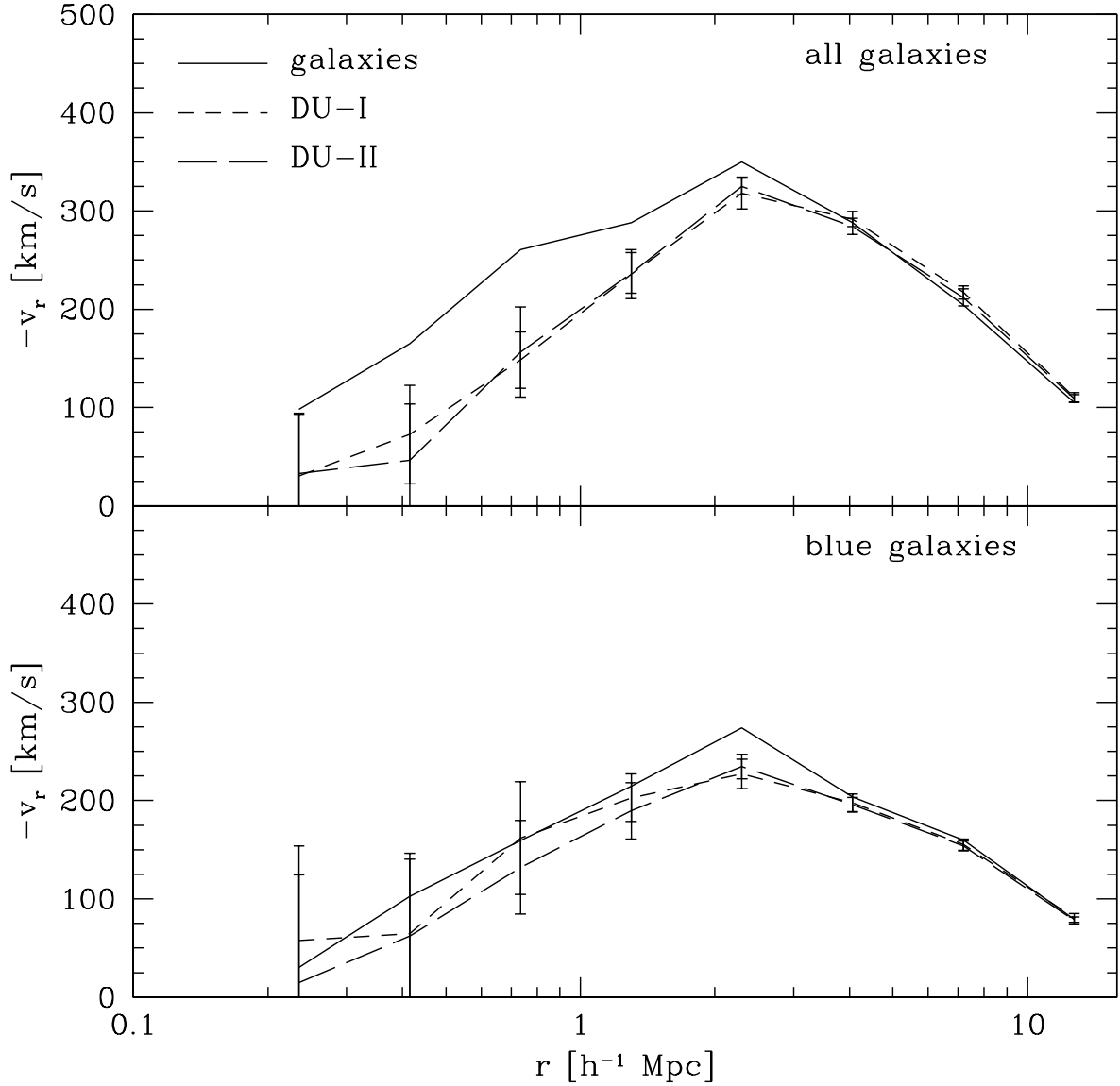}
 \end{center} \figcaption{The mean infall velocity of the DU-I and
 DU-II mock samples of the whole sample, and of the blue galaxies. The
 results from the full simulation for all galaxies and for blue
 galaxies are shown for comparison.\label{fig:mock_vinfall}}
\end{figure}

The mean infall velocity $v_{12}(r)$ of galaxies can be also used to
constrain the cosmological density parameters $\Omega_0$
\citep{Juszkiewicz2000}, according to a simple analytic expression for
$v_{12}$ of the dark matter by \citet{Juszkiewicz1999} as
\begin{equation}
 \label{eq:vinfall}
 v_{12}(r) = -\frac{2}{3}Hr\Omega_0^{0.6}\bar{\xi}(r)[1+\alpha\bar{\xi}(r)],
\end{equation}
\begin{equation}
 \bar{\xi}(r) = \frac{3}{r^3 [1+\xi(r)]}\int_0^r\xi(x)x^2dx,
\end{equation}
where $\alpha=1.2-0.65\gamma$, $\gamma = -(d\ln\xi/d\ln r)_{\xi=1}$ and
$H$ is the Hubble constant. It should be pointed out that in order to
use equation~(\ref{eq:vinfall}) to measure $\beta$, the galaxies used
must trace the underlying dark matter in the spatial and in the velocity
distributions.  Figure~\ref{fig:vinfall} shows $v_{12}(r)$ for dark
matter, all galaxies, and blue and red galaxy samples in our
simulation. Apparently, the blue galaxies have a significantly smaller
infall velocity $v_{12}(r)$ than the red galaxies or all galaxies at all
separation considered. This is not consistent with the result by
\citet{Juszkiewicz2000}, in which they conclude that there is no
difference in $v_{12}$ between early and late-type galaxies. This is
probably because their estimation of $v_{12}(r)$ for ellipticals has
large observational uncertainties. Actually, the number of ellipticals
used in \citet{Juszkiewicz2000} is not sufficient for the definite
estimation of the mean infall velocity. In
Figure~\ref{fig:mock_vinfall}, the mean infall velocity for the DU-I and
DU-II samples of all and of blue galaxies are depicted. The error bars
indicate the statistical dispersion for 20 realizations of the DU-I and
DU-II samples. At large separations with $r\gtrsim 2\mpc$, the mean
infall velocities of the galaxies are well reproduced by the
corresponding DU-I and DU-II samples. Thus, the difference in
$v_{12}(r)$ for all and for blue galaxies is mostly due to the
difference in the spatial distributions. As is done in
\citet{Juszkiewicz2000}, one can determine the values of $\Omega_0$ and
$\sigma_8$ from the value of $v_{12}(r)$ at a certain separation, say
$r=10\mpc$, using equation~(\ref{eq:vinfall}), if the spatial bias of
the observed $v_{12}(r)$ is properly corrected.  However, if one simply
adopts the mean infall velocity of all the galaxies or the blue ones,
which is lower than that of dark matter, the estimated $\Omega_0$ can be
underestimated for a given value of $\sigma_8$ and vice versa
\citep[cf.][]{Juszkiewicz2000}.

\subsection{MASS RECONSTRUCTION ANALYSIS}
In this subsection, we calculate the mass density field reconstructed
using a method similar to the POTENT analysis \citep[][for
review]{Dekel1994, Dekel1999} in order to investigate how the spatial
and dynamical biases on galaxies can affect the results of the POTENT
analysis. In practice, the POTENT analysis attempts to recover the mass
density field from a catalogue of radial velocities of galaxies using
the assumption that the galaxy velocity field is irrotational,
$\nabla\times{\bf v}=0$, and it involves a slightly complicated
statistical treatment of observational data. In this paper, however,
since we are purely interested in the effect of spatial and dynamical
biases, we reconstruct mass density fields from 3-dimensional velocity
data. Specifically, in order to recover the mass density fields, we use
the following quasi-linear approximation of the continuity equation
\citep{Dekel1999}:
\begin{equation}
 \label{eq:POT}
 \delta_{\rm POT} = -(1+\epsilon_1)f^{-1}\nabla\cdot{\bf v}
  -(1+\epsilon_2)f^{-2}\Delta_2+(1+\epsilon_3)f^{-3}\Delta_3,
\end{equation}
where 
\begin{equation}
 \Delta_2=\sum_{i<j}\left\{\left(\fracp{v_i}{x_j}\right)^2
		     -\fracp{v_i}{x_i}\fracp{v_j}{x_j}\right\},
\end{equation}
and
\begin{equation}
 \label{eq:triplet}
 \Delta_3=\sum_{i,j,k}\left(\fracp{v_i}{x_i}\fracp{v_j}{x_k}\fracp{v_k}{x_j}
		     -\fracp{v_1}{x_i}\fracp{v_2}{x_j}\fracp{v_3}{x_k}\right).
\end{equation}
In equation~(\ref{eq:triplet}), the sum is over three cyclic
permutations of $(i,j,k)=(1,2,3)$. The three parameters $\epsilon_1$,
$\epsilon_2$ and $\epsilon_3$ are empirically tuned to best fit the dark
matter density field, and we adopt $\epsilon_1=0.06$,
$\epsilon_2=-0.13$, and $\epsilon_3=-0.3$ throughout this paper.

In the analysis, we calculate the mass and galaxy density fields using
the cloud-in-cloud (CIC) scheme. We also compute a volume weighted
velocity field of galaxies with a smoothing scale $R_s$ according to the
following steps \citep{Babul1994, Berlind2001}. First, we create the
momentum field of galaxies using CIC binning and smooth it with a
Gaussian filter of a smoothing scale $R_1$, where $R_1$ must be much
smaller than the smoothing scale $R_s$ . Second, we divide the smoothed
momentum field by a similarly smoothed density field of galaxies. Note
that this is a mass-weighted velocity field smoothed over the scale
$R_1$. Finally, we smooth this velocity field with another Gaussian
filter of smoothing scale $R_2=(R_s^2-R_1^2)^{1/2}$ to form the
volume-weighted velocity field smoothed with the effective smoothing
scale of $R_s$. In our analysis, we adopt $R_1=1\mpc$ and we calculate
the reconstructed density field $\delta_{\rm POT}$ for $R_s=8\mpc$ and
$12\mpc$ according to equation~(\ref{eq:POT}). The smoothing scale
adopted in the actual POTENT analyses is typically $R_s=10-12\mpc$ in
order to minimize the effect of inhomogeneous Malmquist bias. In this
paper, however, we are interested in the effects of the spatial and
dynamical biases on the POTENT method, and apply it for a slightly
smaller smoothing scale $R_s=8\mpc$.

\begin{figure}[tbp]
 \leavevmode
 \begin{center}
  \plotone{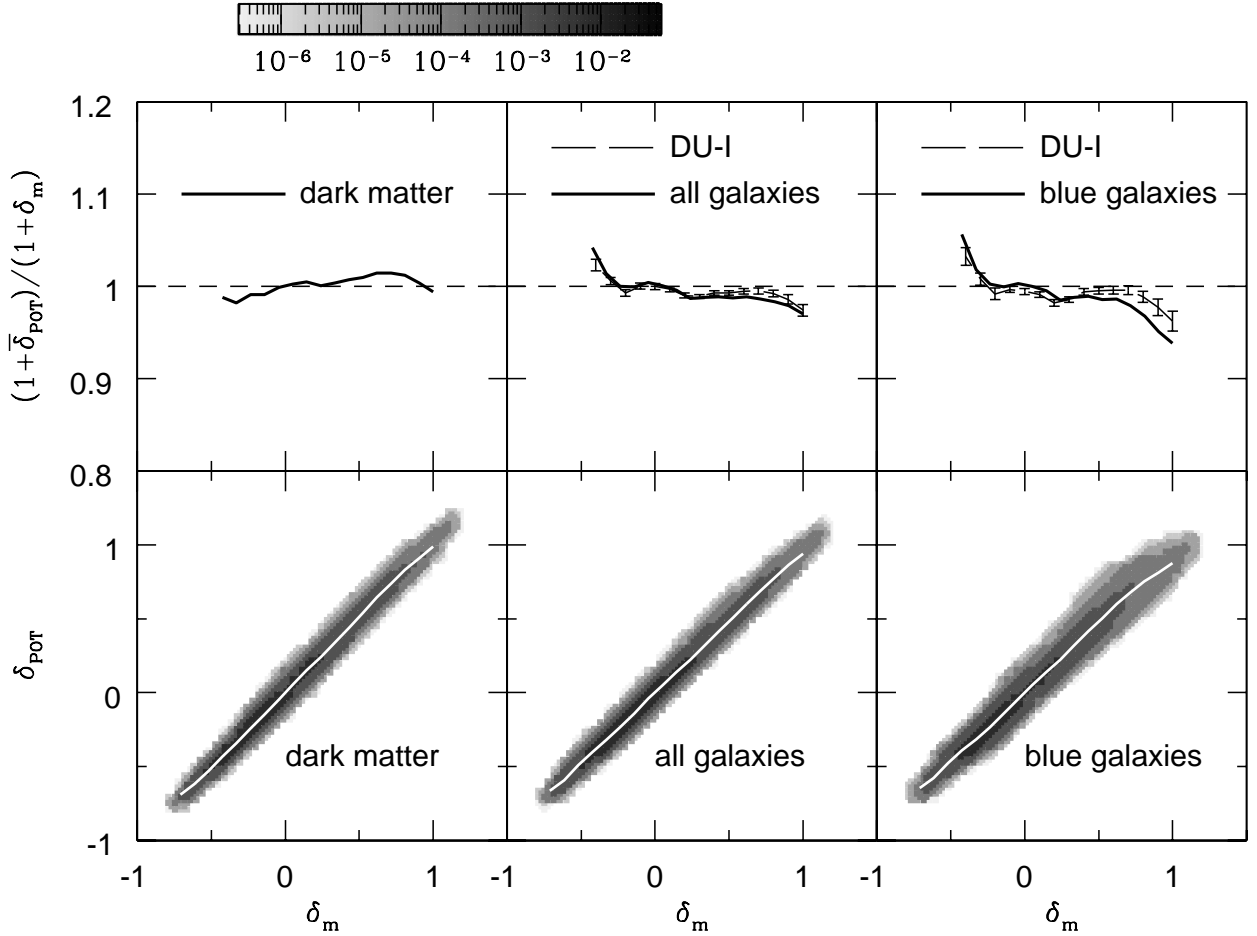} \figcaption{{\it Lower} panels show the joint
  probability distributions of the mass density field $\delta_{\rm m}$
  and that reconstructed using the POTENT-style method $\delta_{\rm
  POT}$. The gray scale indicates the joint probability that a region
  has a set of values $\delta_{\rm m}$ and $\delta_{\rm POT}$,
  simultaneously. The solid lines indicate the mean relation
  $\bar{\delta}_{\rm POT}(\delta_{\rm m})$ between $\delta_{\rm m}$ and
  $\delta_{\rm POT}$. The smoothing scale is set to $R_s=12\mpc$. {\it
  Upper} panels indicate the ratio of $(1+\bar{\delta}_{\rm POT})$ to
  $(1+\delta_{\rm m})$ as a function of $\delta_{\rm m}$. We also show
  the ratio calculated from the DU-I sample in long-dashed lines.
  \label{fig:POTENT12}}
 \end{center}
\end{figure}

\begin{figure}[tbp]
 \leavevmode
 \begin{center}
  \plotone{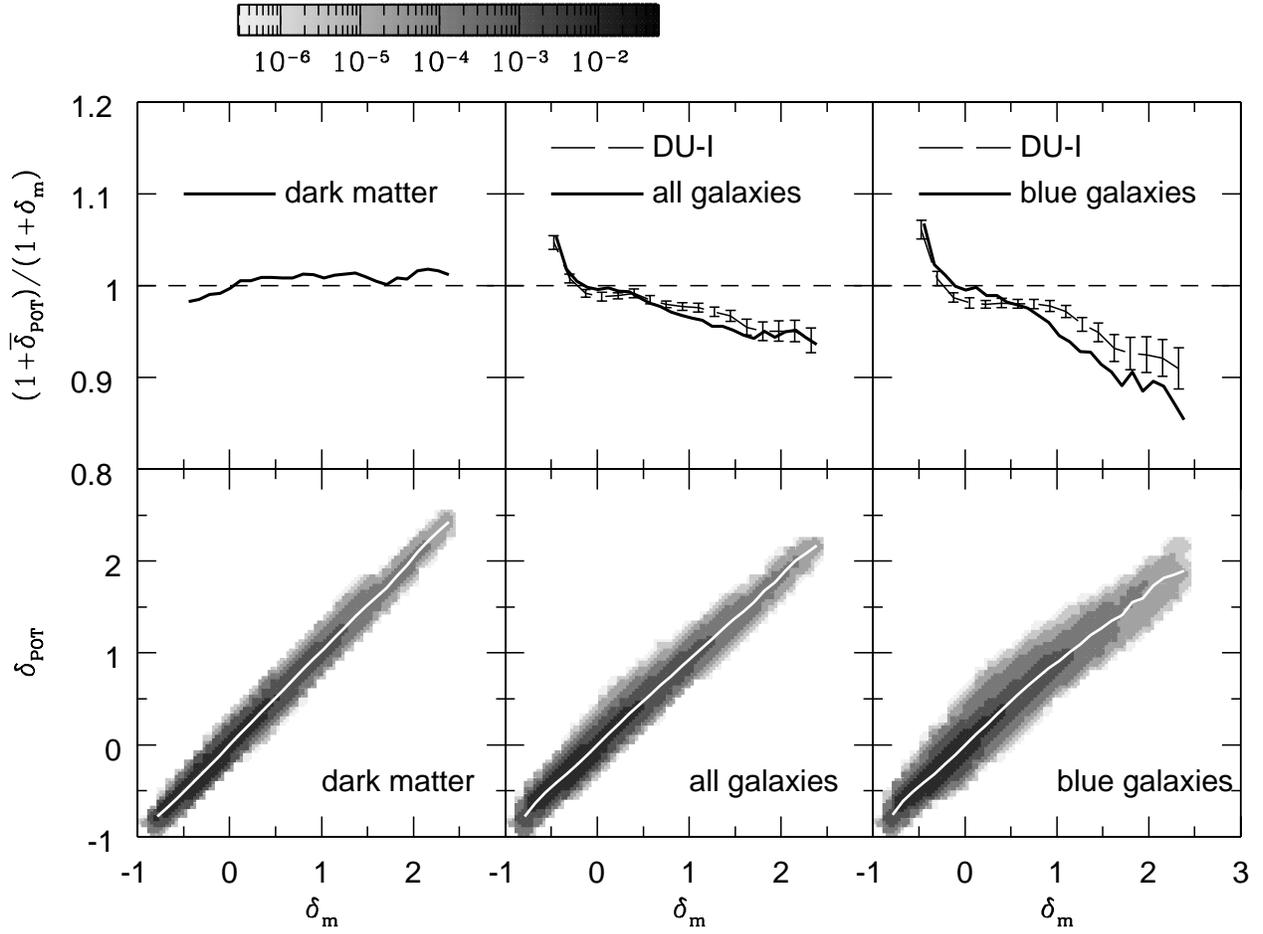}
  \figcaption{Same as Fig.~\ref{fig:POTENT12} but for
  $R_s=8\mpc$. \label{fig:POTENT08}}
 \end{center}
\end{figure}

Figures~\ref{fig:POTENT12} and \ref{fig:POTENT08} show the joint
probability distributions of $\delta_{\rm m}$ and $\delta_{\rm POT}$
reconstructed from the velocity fields of dark matter, all galaxies, and
blue galaxies ({\it lower} panels) for $R_s=12\mpc$ and $R_s=8\mpc$,
respectively. The gray scale indicates the joint probability that a
region has a set of the actual mass overdensity $\delta_{\rm m}$ and the
reconstructed one $\delta_{\rm POT}$, simultaneously. We also show the
ratio of $(1+\bar{\delta}_{\rm POT})$ relative to $(1+\delta_{\rm m})$
in solid lines ({\it upper} panels), where $\bar{\delta}_{\rm POT}$ is
the mean of $\delta_{\rm POT}$ for a given $\delta_{\rm m}$. For both
the smoothing scale of $R_s=8\mpc$ and $12\mpc$, the dark matter
velocity fields well reproduce the real density field with good
accuracy. This indicates that the adopted reconstruction method provides
correct density estimation if a spatially and dynamically unbiased
sample of particles is given. This also confirms that equation
(\ref{eq:POT}) works well for the quasi-linear regions
\citep{Dekel1999}.  In the case of $R_s=12\mpc$, the reconstructed
density fields agree well with the real ones for $\delta_{\rm m}\gtrsim
0$ irrespective of the adopted tracers, while $\delta_{\rm POT}$
reconstructed from the velocity of the blue galaxies is slightly
underestimated by $\simeq 5\%$ at high density ($\delta_{\rm m} \simeq
1$) regions. On the other hand, for $R_s=8\mpc$, the reconstructed
density fields from the velocity fields of the blue and of all the
galaxies are systematically underestimated at high density regions by
$\sim 15\%$ and $\sim 5\%$, respectively.

In order to estimate to what extent the effects of dynamical friction
and spatial bias are important for the discrepancy between $\delta_{\rm
m}$ and $\delta_{\rm POT}$, we carry out the POTENT-style density
reconstruction for the DU-I and DU-II samples of all and of blue
galaxies, and compare the results with those of galaxy samples.  In the
upper panels of Figures~\ref{fig:POTENT12} and \ref{fig:POTENT08},
long-dashed lines with error bars show the ratio of $(1+\delta_{\rm
POT})$ and $(1+\delta_{\rm m})$ reconstructed from the velocity fields
of the DU-I sample. We omit the results from the DU-II sample since they
are almost identical to the DU-I sample. The error bars indicate the
statistical dispersion for 20 realizations of the DU-I samples. In the
case of $R_s=12\mpc$, which is usually adopted in recent POTENT
analyses, we can see that the POTENT-style analyses are not seriously
affected by the spatial and dynamical biases, although the dynamical
bias is only slightly effective at the high density regions for blue
galaxies. In adopting the smoothing scale $R_s=8\mpc$, we have no
significant difference between $\delta_{\rm POT}$ recovered from the
DU-I sample and from all the galaxies, indicating that the dynamical
bias is not important for all galaxies and that only the spatial bias is
responsible for the discrepancy. On the other hand, as for the blue
galaxy samples, although the density field reconstructed from the DU-I
sample of blue galaxies underestimates the real one, the one
reconstructed from the velocity of the blue galaxies is still lower than
that of the DU-I sample. This indicates that the dynamical bias is not
negligible for blue galaxies on this smoothing scale.

As for the spatial bias, as is shown in \citet{Yoshikawa2001}, blue
galaxies in our simulation are anti-biased in high density regions.
Thus, the velocity field calculated from the blue galaxies is
inhomogeneously sampled and does not faithfully represent the underlying
dark matter velocity, especially at high mass density regions and for
smaller smoothing scales. In our analyses, in the case of $R_s=8\mpc$,
as shown in Figure~\ref{fig:POTENT08}, the effect of the spatial bias
leads to an underestimate of the real density field by $\simeq
10\%$. Actually, much work has been done to understand and control this
problem \citep{Dekel1999}. Other reconstruction methods and some
algorithms to estimate $\beta$ parameter from velocity data of galaxies
have been proposed \citep{Zaroubi1995, Willick1997, Zaroubi1999} which
prevent the effect of the spatial bias to some extent. However, these
methods still cannot escape from the dynamical bias, because it is
implicitly assumed that there is no dynamical bias in the observed
velocity field.  Therefore, when the velocity data of late-type galaxies
smoothed over the scale of $R_s=8\mpc$ or smaller are adopted, the
reconstructed density field and the estimated $\beta$ value can be
skewed due to the dynamical bias.

\subsection{MOTION OF GALAXIES WITHIN DARK HALOS}
\label{ss:motion_of_galaxies}

\begin{figure}[tbp]
 \leavevmode
 \begin{center}
  \plotone{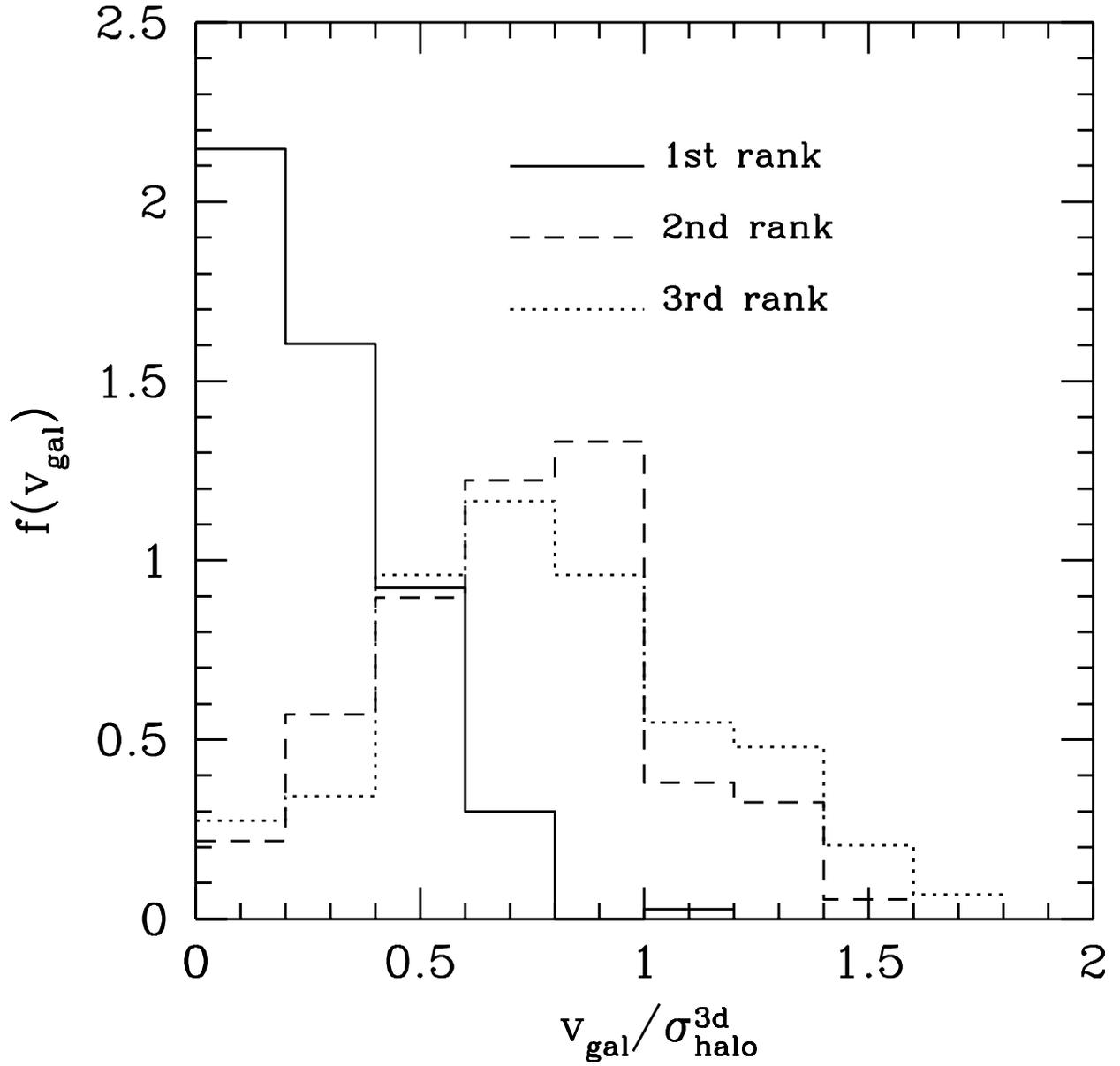}
  \figcaption{Distributions of relative velocity of the three most
  massive galaxies to each host dark halo. The velocity of galaxies
  is scaled by the 3-dimensional velocity dispersion of the dark
  matter particles inside their host halos. \label{fig:vbias}}
 \end{center}
\end{figure}

\begin{figure}[tbp]
 \leavevmode 
 \begin{center}
  \plotone{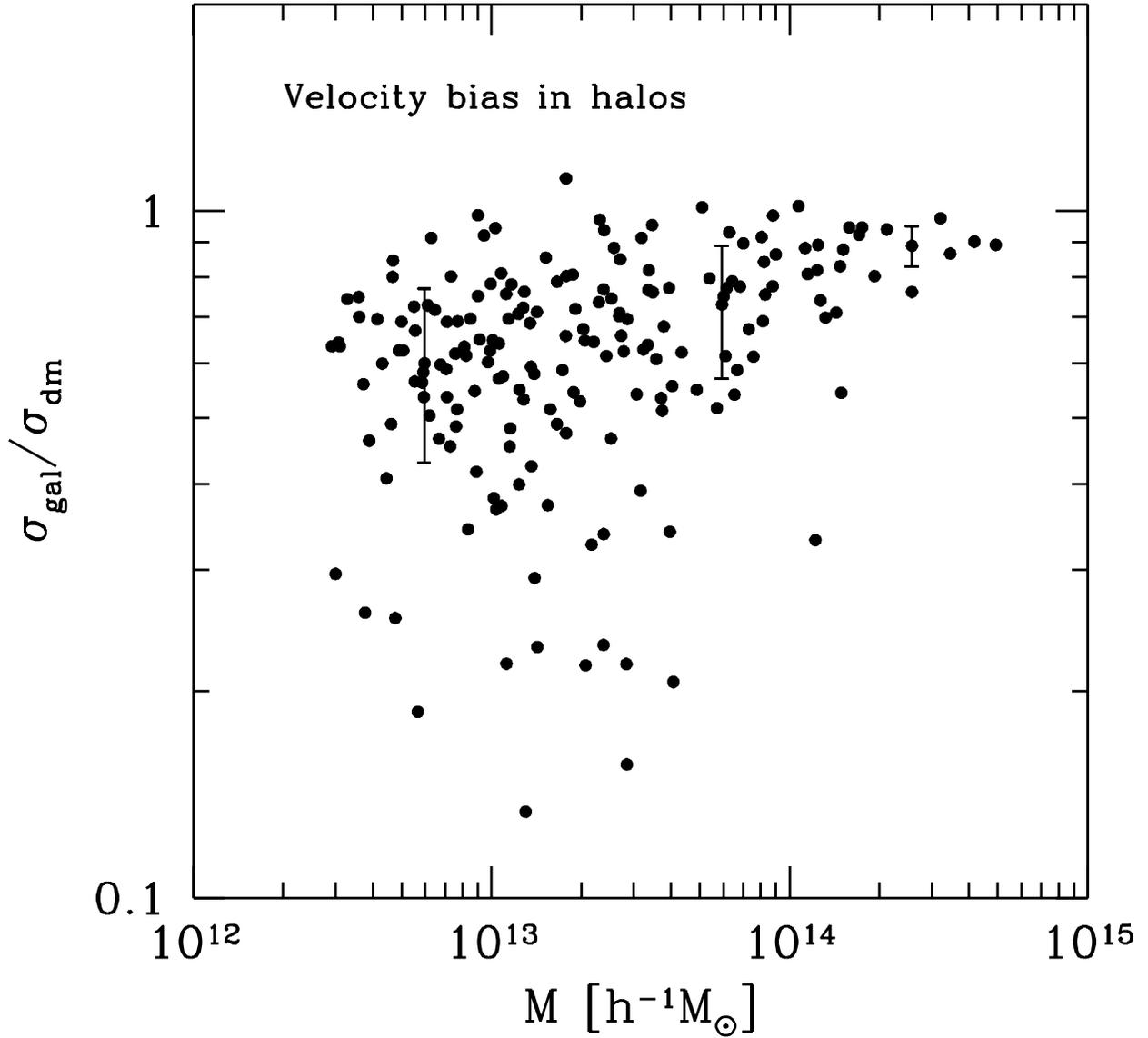} \figcaption{The ratio of the velocity dispersion
 of galaxies to that of dark matter in each dark halo as a function of
 halo mass. Representative error bars for individual halos of
 different mass are also indicated. \label{fig:halo_vbias}}
 \end{center}
\end{figure}

So far, we have shown the velocity statistics of galaxies in a
cosmological volume. In this subsection, we focus on the statistics of
motion of galaxies relative to their host dark halos.
Figure~\ref{fig:vbias} shows the distributions of velocities of the
three most massive galaxies in each dark halo relative to the average
velocity (i.e. the velocity of the mass center) of the host dark
halo. Here, we normalized the velocity of the galaxies by the
3-dimensional velocity dispersion of the dark matter inside the dark
halos. The most massive galaxy in each dark halo is almost static
within the host dark halo, consistent with the observation of cD
galaxies in galaxy clusters \citep{Oegerle01}. The distributions of
the second and the third ranked galaxies are skewed toward the smaller
velocities, because they should have a peak at $v_{\rm
gal}/\sigma^{\rm 3d}_{\rm halo}\simeq 1$ if the motion of the galaxies
is the same as that of the dark matter, free from velocity bias.
Furthermore, the ratio of the velocity dispersions of galaxies
$\sigma_{\rm gal}$ and dark matter $\sigma_{\rm dm}$ inside each dark
halo are plotted in Figure~\ref{fig:halo_vbias}. Representative
error bars for halos of different mass are indicated in the figure that
are the standard scatters of the dispersion ratio at that halo
mass. Velocity dispersions of galaxies in relatively low mass
($\simeq 10^{13} M_{\odot}$) dark halos are systematically smaller
than those of dark matter, while for very massive dark halo with mass
greater than $10^{14}M_{\odot}$, we have $\sigma_{\rm gal} \simeq
\sigma_{\rm dm}$. This result is due to the facts that the central
galaxies are static, and the other galaxies are slowed down,
significantly especially in small halos, by dynamical friction. This
result also shows why the PVDs of blue galaxies are small, for the
blue galaxies mainly reside in smaller dark halos, and the PVDs at
small separation have their main contribution from pairs of galaxies
within the same dark halo. If we use a larger mass cut for
defining galaxies, we expect the dispersion ratio will systematically
decrease, because the dynamical friction bias is more effective for
more massive objects (see \S 4 for a further discussion). 

From these two statistics, we find that in each dark halo the motion of
galaxies is affected by dynamical friction, and their velocity is less
than that of dark matter. Since the blue galaxies in our simulation
statistically tend to be less massive and preferentially reside in
smaller dark halos, these results indicate that the blue galaxies are
selectively affected by dynamical friction. Furthermore, the implication
for the velocity dispersion of galaxies, which has been conventionally
used to estimate the mass of clusters and groups of galaxies
\citep{Bahcall1981, Zabludoff93, Wu97, Girardi2000, Girardi2002}, is
that the mass of groups of galaxies based on the velocity dispersion of
galaxies could be seriously underestimated.

\section{SUMMARY AND DISCUSSION}
In this paper, we have investigated several velocity statistics of
galaxies and their biases relative to dark matter particles using
cosmological hydrodynamic simulations. We consider the PVD, the mean
infall velocity, the POTENT-style analysis for all galaxies identified
in our simulation and for two subsamples, the blue and red galaxies,
categorized based on their formation epochs. We also construct
dynamically unbiased samples of all and of blue galaxies, which mimic
the spatial distributions of galaxies but still keep the same dynamical
properties as the dark matter, and measure their velocity statistics in
order to investigate the effects of the spatial and the dynamical biases
of galaxies on their velocity statistics.

Our major findings are as follows:

1. The PVD of all galaxies is systematically lower than that of the dark
matter, and the blue galaxies have a lower PVD than galaxies from the
whole sample, especially at small separation. This is consistent with
previous observational results \citep{Jing1998, Jing2002,
Zehavi2002}. According to our analyses, the low PVD of all the galaxies
is due mostly to the difference of the spatial distributions between
galaxies and dark matter. On the other hand, the discrepancy in the PVD
between dark matter and the blue galaxies cannot be explained only by
the spatial bias, and it is suggested that there must be an effect of
the dynamical bias for the blue galaxies.

2. The mean infall velocity of the blue galaxies is significantly lower
than that of the whole sample, and of the red galaxies. The dark matter
particles show almost the same values as the red galaxies at $r\gtrsim
3\mpc$ and $\simeq 100\kms$ lower than the whole sample and the red
galaxies at $r \lesssim 1\mpc$.  In estimating $\Omega_0$ and $\sigma_8$
from the mean infall velocity of all the galaxies or of the late-type
galaxies at the separation of $r=10\mpc$, one will underestimate
$\Omega_0$ for a given value of $\sigma_8$ and vice versa. We also find
that the difference at the separation $r\gtrsim 3\mpc$ is due only to
the difference in the spatial distribution among galaxy populations.

3. Our analysis shows that the POTENT-style mass density reconstruction
using galaxies' peculiar velocity is almost free from the spatial and
dynamical biases on the smoothing scale of $R_s=12\mpc$, which is
typically adopted in recent POTENT analyses. On the other hand, in the
case of $R_s=8\mpc$, it suffers from spatial bias, if the velocity data
of all galaxies are adopted, and furthermore, the dynamical bias is also
important when we use the velocity field of the blue galaxies. Some
methods to probe the $\beta$ parameter based on velocity-velocity
comparisons \citep{Davis1996, Willick1997}, and the mass density
reconstruction based on the Wiener filtering \citep{Zaroubi1995,
Zaroubi1999} can evade the effect of the spatial bias to some extent,
and probe the mass density or velocity on smaller scales, but still they
can suffer from the dynamical bias on smoothing scales equal to or
smaller than $R_s=8\mpc$, if the velocity data of late-type galaxies are
adopted.

4. In each dark halo, the most massive galaxy is almost static within
the dark halo. The velocity of the second and third ranking massive
galaxies relative to their host dark halos is statistically smaller
than that of the dark matter inside the halos due to dynamical
friction. Furthermore, the velocity dispersions of galaxies in less
massive ($\sim 10^{13}M_{\odot}$) dark halos are systematically lower
than those of dark matter particles inside dark halos. This bias in the
velocity dispersion of the galaxies leads to serious underestimates of
the mass of groups of galaxies.  It is also responsible for the low PVD
value of the blue galaxies.

The reason why the blue galaxies are selectively affected by dynamical
friction is tightly related to their formation epochs and can be
qualitatively explained as follows. According to the standard scenario
of galaxy formation, galaxies form in the gravitational potential wells
provided by the dark matter halos and dark matter clumps inside dark
halos. The red galaxies are formed within more massive halos than the
blue ones, and they experience a significant number of merging
processes. The mergers mix gas and dark matter, and thus reduce the
differences in the distribution. The blue galaxies form in less massive
halos, and have less mergers during their formation history. Their mass
is a bigger fraction of the host halo mass, and thus they are
efficiently slowed down relative to the center of mass of the halo by
dynamical friction.

Our definition of galaxies in our simulation data is admittedly rather
phenomenological. A more observationally oriented classification of
galaxies, for example using color or magnitude of galaxies, is needed
for a direct comparison with observations. We plan to implement more
realistic prescriptions of galaxy formation and evolution including star
formation, feedback and UV background heating in due course. However, it
is encouraging that even these simple prescription of galaxy formation
can reproduce the clustering and the PVDs of the observed galaxies. Thus
we believe that the results presented in this paper are quite robust and
that it is worth considering the importance of dynamical friction for
the velocity statistics of galaxies.

\acknowledgments

We thank Yaushi Suto and Takashi Hamana for several enlightening
discussions, and an anonymous referee for suggesting some improvements
in this manuscript. Numerical simulations presented in this paper were
carried out at ADAC (the Astronomical Data Analysis Center) of the
National Astronomical Observatory, Japan, and at KEK (High Energy
Accelerator Research Organization, Japan).  Y.P.J. was supported in part
by NKBRSF (G19990754) and by NSFC (No.10125314).

\newpage



\end{document}